# Tunable stress and controlled thickness modification in graphene by annealing

*Zhen Hua Ni,[†, §] Hao Min Wang,[‡] Yun Ma,[†] Johnson Kasim,[†] Yi Hong Wu,[‡] and Ze Xiang Shen,[†,*]*

[†]Division of Physics and Applied Physics, School of Physical and Mathematical Sciences, Nanyang Technological University, 1 Nanyang Walk, Block 5, Level 3, Singapore 637616
[‡]Department of Electrical and Computer Engineering, National University of Singapore, 4 Engineering Drive 3, Singapore 117576
[§]Department of Physics, National University of Singapore，2 Science Drive 3, Singapore 117542

**ABSTRACT**

Graphene has many unique properties which make it an attractive material for fundamental study as well as for potential applications. In this paper, we report the first experimental study of process-induced defects and stress in graphene using Raman spectroscopy and imaging. While defects lead to the observation of defect-related Raman bands, stress causes shift in phonon frequency. A compressive stress (as high as 2.1 GPa) was induced in graphene by depositing a 5 nm $SiO_2$ followed by annealing, whereas a tensile stress (~ 0.7 GPa) was obtained by depositing a thin silicon capping layer. In the former case, both the magnitude of the compressive stress and number of graphene layers can be controlled or modified by the annealing temperature. As both the stress and thickness affect the physical properties of graphene, this study may open up the possibility of utilizing thickness and stress engineering to improve the performance of graphene-based devices. Local heating techniques may be used to either induce the stress or reduce the thickness selectively.

Graphene, a monolayer graphite sheet, has attracted much interest since it was discovered in 2004.[1-3] The exceptionally high crystallization and unique electronic properties make graphene a promising candidate for ultrahigh speed nanoelectronics.[4] However, in order to make it a real technology, several critical issues need to be resolved which include but are not limited to (1) microelectronics compatible processes for fabricating both single layer and few layer graphene and related devices and (2) viable way of creating an energy gap at K and K′ points in the Brillouin zone. Researchers have successfully developed an energy gap in graphene by patterning it into nanoribbon,[5] forming quantum dots [4] or making use of mutilayer graphene sheets with or without the application of an external electrical field.[6,7] Besides global back gates,[8,9] top local gates[10-12] have also been employed to develop more complex graphene devices, such as pn junction,[13] Veselago lens[14] and Klein tunneling.[15] The top gate oxides that have been used so far include $H_fO_2$, $Al_2O_3$ and $SiO_2$. Although efforts have been made to deposit the gate oxides without damaging the graphene or changing its electrical properties,[10-15] the gate oxides should influence the graphene sheets in at least three ways: doping, defects, and various mechanical deformations. Although theoretical studies suggest that chemical doping shifts the neutral point [1,16,17] and defects increase carrier scattering in graphene,[18,19] so far they have not been studied experimentally. It is known that the $sp^2$ bonds in graphitic carbon can hold extremely high mechanical strains [20] and exhibit interesting electromechanical properties, as observed in carbon nanotubes (CNTs).[21] Remarkable strain/stress effects on optical and electronic properties have been found in CNTs.[21-25] As both the CNTs and graphene share the same honeycomb structure,[26] it is plausible to expect similar type of effects in graphene, especially in gapped structures such as graphene nanoribbon, quantum dot, and nano-constrictions.

We have studied systematically graphene sheets subjected to defects and mechanical deformations induced by insulating capping layers using Raman spectroscopy and Raman microscopy. Different insulating materials were deposited on top of graphene by electron beam evaporation, pulsed laser deposition (PLD), sputtering and followed by annealing at different temperatures. Here we present the results of using $SiO_2$ as a typical example. Thin layer of $SiO_2$ (5 nm) was deposited on top of the graphene sheets by PLD and Raman spectroscopy was used to investigate the interaction between the $SiO_2$ and graphene. Defect-induced Raman bands were observed after the deposition of $SiO_2$. The amount of defects was significantly reduced by annealing. A striking feature in our spectroscopic data is that compressive stress as high as ~2.1 GPa was observed after annealing process. The compressive stress may be useful to tune the electronic properties of graphene nanostructures. Possible applications to graphene based devices and spectroscopic research are also presented. To the best of our knowledge, this is the first experimental report on defects and stress induced in graphene. We further show that the graphene thickness, and hence its properties, can be changed in a controlled manner by annealing in air.

**RESULTS AND DISCUSSION**

Figure 1a shows the optical image of a graphene sample on the $SiO_2$/Si substrate. The graphene sheet shows four different contrast regions, which can be attributed to four different thicknesses. The Raman spectra recorded from these regions are shown in Figure 1b. There are two intense features in the spectra, which are the in-plane vibrational ($E_{2g}$) G band and the two phonon 2D band, respectively. As has been proposed by Ferrari et al.[27] the second order Raman 2D band is sensitive to the number of layers of graphene and the 2D band of single layer graphene is very sharp and symmetric. In our Raman spectra, the sharp 2D band of the single layer graphene can be clearly observed and distinguished from bilayer and few-layer graphenes. We can further identify the thickness of other layers from the G band intensity plot, as shown in Figure 1c, since the intensity of G band increases almost linearly with the number of layers for few-layer graphene samples.[28] Figure 1d plots the

Raman intensity of the G band along three dash lines drawn in Figure 1c. It is obvious that the graphene sheet contains one, two, three and four layers.

The Raman spectra of graphene before and after 5 nm SiO$_2$ deposition were shown in Figure 2a. A clear difference is that two extra Raman bands, located at 1350 and 1620 cm$^{-1}$, were observed after deposition. Those two Raman bands were both defects induced: The stronger one at 1350 cm$^{-1}$ is assigned to the so-called disorder-induced D band, which is activated by a double resonance effect by defects, such as in-plane substitutional hetero-atoms, vacancies, or grain boundaries.[29] The weaker band at 1620 cm$^{-1}$ is assigned to D′ band. The D′ band corresponds to the highest frequency feature in the density of state, which is forbidden under defect-free conditions.[30] Its observation is also associated with the presence of defects in the lattice and originates from the double resonance process. The observation of D and D′ bands indicate that defects were introduced into graphene after the 5 nm SiO$_2$ top layer deposition. This may be caused by the damage on the sample during deposition, or by the interaction between SiO$_2$ and graphene which may produce vacancy, dislocation and/or dangling bonds. The defect peaks were also observed in graphene with 5 nm SiO$_2$ top layer deposited by e-beam evaporation. Annealing is carried out to eliminate the defects, which will be discussed in latter section. Figure 2b shows the Raman spectra of graphene sheet with one to four layers as well as that of bulk graphite after SiO$_2$ deposition. The Raman spectra were taken under same conditions. The D band intensity decreases with the increase of graphene thickness and is invisible for bulk graphite, demonstrating that defects are more easily introduced into thinner graphene sheets.[31] Figure 2c and 2e show Raman images generated from the intensity of D band before and after deposition respectively. Before deposition, there is no D band hence the Raman image is dark. After deposition, the thinner graphene (single layer graphene) shows the strongest D band, which is consistent with the discussion above. Figure 3d and 3f show the images generated from the intensity of the corresponding G band, and they do not show noticeable difference. Hence the G band intensity is still a good criterion in determining the thickness of the graphene sheet.

We have also deposited different materials as capping layers with different methods as shown in Fig 3. $SiO_2$ layers were deposited with different methods: electron beam evaporation, PLD and RF sputtering. Different amounts of defects were introduced into the graphene sheets, as indicated by the relative intensity of the defect-induced D band. After $HfO_2$ thin layer deposition by PLD, strong defect-induced D band was observed. However, after polymethyl methacrylate (PMMA) deposition by spin coating, no change in Raman features was observed, as shown in Fig. 3. Our results show that the deposition methods have a significantly effect on the defects, with spin coating introducing the least amount of defects and PLD and RF sputtering the most defects.

The Raman spectra of single layer graphene after annealing in air ambient at different temperatures are shown in Figure 4a. We have also carried out vacuum annealing and similar results were observed. An obvious observation is that the intensity of D band decreases upon annealing. This is clearly demonstrated in Figure 4b, which shows the intensity ratio between the D band and G band ($I_D/I_G$) that is often used to estimate the amount of defects in carbon materials. For one to four-layer graphene sheets, this ratio decreases with increase in annealing temperature. This can be understood as due to the recovery of damaged graphene at high temperature. Figure 5a-c show another important observation, where the G, D, and 2D bands shifted to higher frequency with increase in annealing temperature. The G band blue shifted ~15 $cm^{-1}$, while the D band blue shifted ~13 $cm^{-1}$ and 2D band ~25 $cm^{-1}$ after annealing at 500 $^oC$. We attribute this significant blueshift of Raman bands to the strong compressive stress on graphene. The $SiO_2$ becomes denser upon anneal so it exerted a strong compressive stress on the graphene. For comparison, the Raman bands of bulk graphite did not shift after deposition and annealing, which supported the above explanation, as bulk graphite is too thick and it is not easily compressed by SiO$_2$. Recently, Yan et al.[32] and Pisana et al.[33] found that the frequency of the G and 2D Raman bands can also be adjusted by charge doping through electron-phonon coupling change. Besides the G band blueshift, a bandwidth narrowing of ~10 cm$^{-1}$ was also observed in the case of charge doping. However, in our results, only a small fluctuation (±1 cm$^{-1}$) of G band FWHM (full width at half maximum) was observed after annealing at different temperature, which indicates that the effect of charge doping can be ignored. In addition, it is shown that the dependence of the 2D band blueshift on doping is very weak and only ~10-30% compared to that of G band.[32,34] Hence, the 25 cm$^{-1}$ 2D band blueshift is too large to be achieved by charge doping alone. Therefore, the observed shifts of G (~15 cm$^{-1}$) and 2D (~25 cm$^{-1}$) band in our experiment were mainly caused by stress.

The compressive stress on graphene in our experiment is due to the denser of SiO$_2$ upon annealing. This origin of the compressive stress is very similar to the biaxial stress due to the lattice mismatch at the sample/substrate interface in a normal thin film. Therefore, the stress on graphene should be biaxial. The biaxial compressive stress on graphene can be estimated from the shift of Raman E$_{2g}$ phonon with the following analysis.

For a hexagonal system, the strain ε induced by an biaxial stress σ can be expressed as:[35,36]

$$\begin{bmatrix} \varepsilon_{xx} \\ \varepsilon_{yy} \\ \varepsilon_{zz} \\ \varepsilon_{yz} \\ \varepsilon_{zx} \\ \varepsilon_{xy} \end{bmatrix} = \begin{bmatrix} S_{11} & S_{12} & S_{13} & & & \\ S_{12} & S_{11} & S_{13} & & & \\ S_{13} & S_{13} & S_{33} & & & \\ & & & S_{44} & & \\ & & & & S_{44} & \\ & & & & & 2(S_{11}-S_{12}) \end{bmatrix} \begin{bmatrix} \sigma \\ \sigma \\ 0 \\ 0 \\ 0 \\ 0 \end{bmatrix}, \quad (1)$$

with the coordinate $x$ and $y$ in the graphite/graphene plane and $z$ perpendicular to the plane. So that: $\varepsilon_{xx} = \varepsilon_{yy} = (S_{11}+S_{12})\sigma$, $\varepsilon_{zz} = 2S_{13}\sigma$, $\varepsilon_{yz} = \varepsilon_{zx} = \varepsilon_{xy} = 0$.

With all shear components of strain equal to zero, the secular equation of such system can be written as:

$$\begin{vmatrix} A(\varepsilon_{xx} + \varepsilon_{yy}) - \lambda & B(\varepsilon_{xx} - \varepsilon_{yy}) \\ B(\varepsilon_{xx} - \varepsilon_{yy}) & A(\varepsilon_{xx} + \varepsilon_{yy}) - \lambda \end{vmatrix} = 0, \qquad (2)$$

where $\lambda = \omega_\sigma^2 - \omega_0^2$, with $\omega_\sigma$ and $\omega_0$ the frequencies of Raman $E_{2g}$ phonon under stressed and unstressed conditions.

There is only one solution for this quation:

$$\lambda = A(\varepsilon_{xx} + \varepsilon_{yy}) = 2A\varepsilon_{xx} = 2A(S_{11} + S_{12})\sigma. \qquad (3)$$

Therefore,

$$\omega_\sigma - \omega_0 = \frac{\lambda}{\omega_\tau + \omega_0} \approx \frac{\lambda}{2\omega_0} = \frac{A(S_{11} + S_{12})\sigma}{\omega_0} = \alpha\sigma, \qquad (4)$$

where $\alpha = \dfrac{A(S_{11} + S_{12})}{\omega_0}$ is the stress coefficient for Raman shift.

Using A= -1.44 × 10$^7$ cm$^{-2}$ [35] and graphite elastic constants $S_{11}$=0.98 × 10$^{-12}$ Pa$^{-1}$ and $S_{12}$= -0.16 × 10$^{-12}$ Pa$^{-1}$,[37] and $\omega_0$ =1580 cm$^{-1}$, the stress coefficient $\alpha$ is estimated to be 7.47 cm$^{-1}$/GPa. The estimated stress on single layer graphene with annealing temperature is shown in Figure 5d. The compressive stress on graphene was as high as ~2.1 GPa after depositing SiO$_2$ and annealing at 500 $^o$C, and the stress on single layer graphene in our experiment can be fitted by the following formula:

σ = -0.155 +2.36×10$^{-3}$ T+5.17×10$^{-6}$ T$^2$ (5)

where σ is the compressive stress in GPa and T is temperature in °C. The appearance of such large stress is mainly because graphene sheets are very thin (0.325 nm in thickness for single layer graphene),[38] so that they can be easily compressed or expanded. It has been reported that even the very weak van der waals interaction can produce large stress on the single wall carbon nanotubes.[25] We have also introduced tensile stress onto graphene by depositing a thin cover layer of silicon. The G band of graphene red shifted by ~5 cm$^{-1}$ after silicon deposition, which corresponds to a tensile stress of ~0.67 GPa on graphene sheet. We suggest that tensile stress can be also achieved by depositing other materials with larger lattice constant than graphene. In combination with annealing, both compressive and tensile stress can be introduced and modified in graphene in a controllable manner. The stressed graphene may have very important applications as the properties of graphene (optical and electronic properties) can be adjusted by stress, where stress studies in CNTs have already set good examples,[21-25] e.g. the bandgap of CNTs can be tuned by strain with a parameter of 100 meV per 1% strain.[22] Stress engineering using SiGe alloy has already been used in the IC fabrication to improve the device performance.

Figure 6a shows the optical image of a graphene sheet with one, two, three, four, five and six-layer regions, as denoted by the numbers on the image. After annealing at 600 °C for 30 min, the thinner part of graphene sheet (one to three layers) disappeared due to oxidation. However, the thicker part (four to six layers) still remained, and the thicknesses were reduced to two, three, and four layers, as shown in Figure 6b. The thickness of different regions before and after anneal is determined by a combination of Raman imaging (Figure 6c and 6d) and contrast imaging (Figure 6e and 6f).[39] Optical spectroscopic imaging techniques have a clear advantage in this case over other techniques, e.g. atomic force microscopy (AFM), in determining the layer thickness, as AFM does not work properly due to the presence of SiO$_2$ top layer on the graphene. Although the exact mechanism of graphene annihilation is unknown, it is most likely due to oxidation of carbon by oxygen diffused through the

SiO$_2$ cover layer from the air ambient as the thickness of the graphenes does not change when anneal is carried out in vacuum. Figure 7 shows the Raman spectra of the remained two and four layers graphene. The D band in both spectra is very weak, indicting the high quality of graphene sheets after thickness modification. This result suggests that annealing in the presence of oxygen provides a practical method of manipulating the graphene thickness in a controllable manner. For example, a local heating techniques may be used to either induce the stress or reduce the thickness selectively, opening another avenue for fabricating graphene-based devices.

**CONCLUSION**

In summary, we have used Raman spectroscopy and microscopy to investigate the influence of top gate insulator (5 nm SiO$_2$) on graphene sheets mainly on two important aspects, defects and stress. The results show that defects were introduced in graphene sheets during deposition and the amounts of defects increase as the graphene thickness decreases. After annealing, the defects in graphene can be greatly reduced. Moreover, significant Raman shifts of all the graphene bands were observed after annealing, which was attributed to the compressive stress on graphene. Importantly, the stress can be controlled by the annealing temperature, which maybe used to tune the optical and electronic properties similar to what has been observed in CNTs. Finally, the graphene thickness can be modified in a controllable manner using anneal. Our findings provide useful information critical to graphene device engineering and fabrication.

**EXPERIMENTAL SECTION**

The graphene samples were prepared by micromechanical cleavage and transferred to Si wafer substrate with a 300 nm $SiO_2$ cap layer.[1] Optical microscopy was used to locate the graphene sheet and the thickness was further confirmed by contrast[39] and Raman spectra/image. A 5 nm $SiO_2$ top layer was deposited by PLD with a 248 nm KrF pulsed laser. The laser power used was very weak (~200 mJ and repetition rate of 10Hz) to achieve the slow and smooth deposition (1Å/min) and ellipsometry was used to measure the total thickness of $SiO_2$. The $SiO_2$ thickness on the Si substrate was 303.5 ± 0.5 nm before deposition and 308.5 ± 0.5 nm after deposition, indicating that the thickness of top $SiO_2$ layer was 5 nm. The sample was annealed in a tube furnace at different temperatures for 30 min.

The Raman spectra were recorded with a WITEC CRM200 Raman system with a double-frequency Nd:YAG laser (532 nm) as excitation source. The laser power at sample is below 0.1 mW to avoid laser induced heating. The contrast of graphene are obtained by the following calculation: $C(\lambda) = (R_0(\lambda) - R(\lambda))/R_0(\lambda)$, where $R_0(\lambda)$ is the reflection spectrum from the $SiO_2$/Si substrate and $R(\lambda)$ is the reflection spectrum from graphene sheet, which is illuminated by normal white light.[39] For Raman/contrast image, the sample was placed on an *x-y* piezostage and scanned under the illumination of laser/white light. The Raman/reflection spectra from every spot of the sample were recorded. The stage movement and data acquisition were controlled using ScanCtrl Spectroscopy Plus software from WITec GmbH, Germany. Data analysis was done using WITec Project software.  A 100× objective lens with a NA=0.95 was used both in the Raman and reflection experiments, and the spot size of 532 nm laser and white light were estimated to be 500 nm and 1 μm, respectively.

Fig. 1

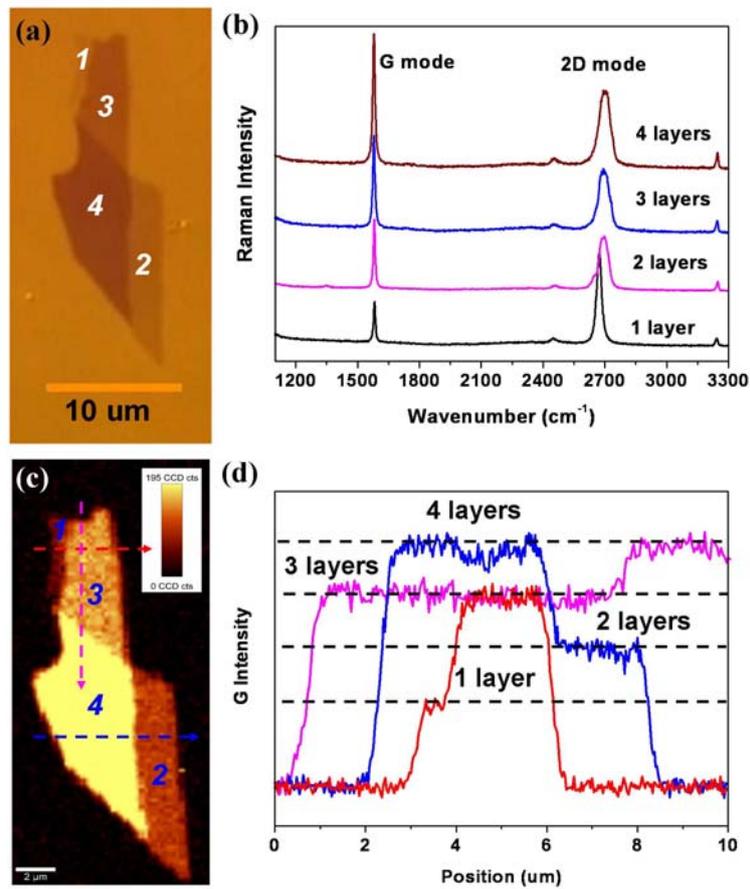

Figure 1. (a) Optical image of graphene with 1, 2, 3 and 4 layers. (b) Raman spectra as a function of number of layers. (c) Raman image plotted by the intensity of G band. (d) The cross section of Raman image, which corresponds to the dash lines with corresponding colors in Raman image. It is obvious that the graphene sheet contains one, two, three and four layers.

Fig.2

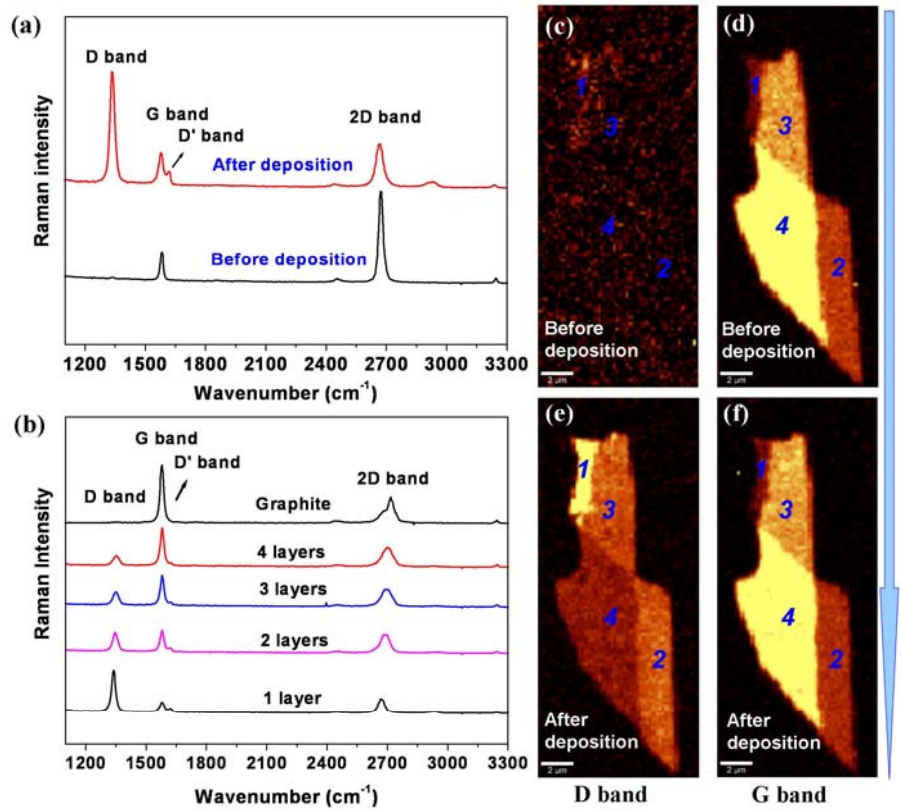

Figure 2. (a) Raman spectra of single layer graphene before and after the 5 nm $SiO_2$ deposition. (b) Raman spectra of graphene with different thicknesses as well as that of bulk graphite after 5 nm $SiO_2$ deposition. Raman images of graphene sheets before $SiO_2$ deposition generated from the intensity of the D band (c) and G band (d). Raman images of graphene sheets after 5 nm $SiO_2$ deposition using the intensity of D band (e), and G band (f). The thinner graphene sheets have stronger D band, hence they contains more defects.

Fig. 3

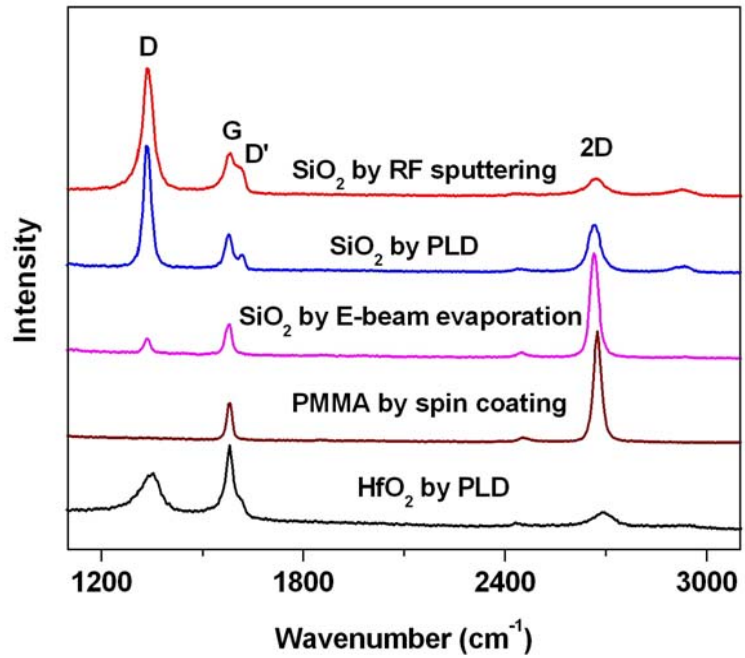

Figure 3. Raman spectra of graphene after SiO$_2$ deposition by RF sputtering, PLD, e-beam evaporation, as well as graphene after PMMA deposition by spin coating and HfO$_2$ deposition by PLD.

Fig. 4

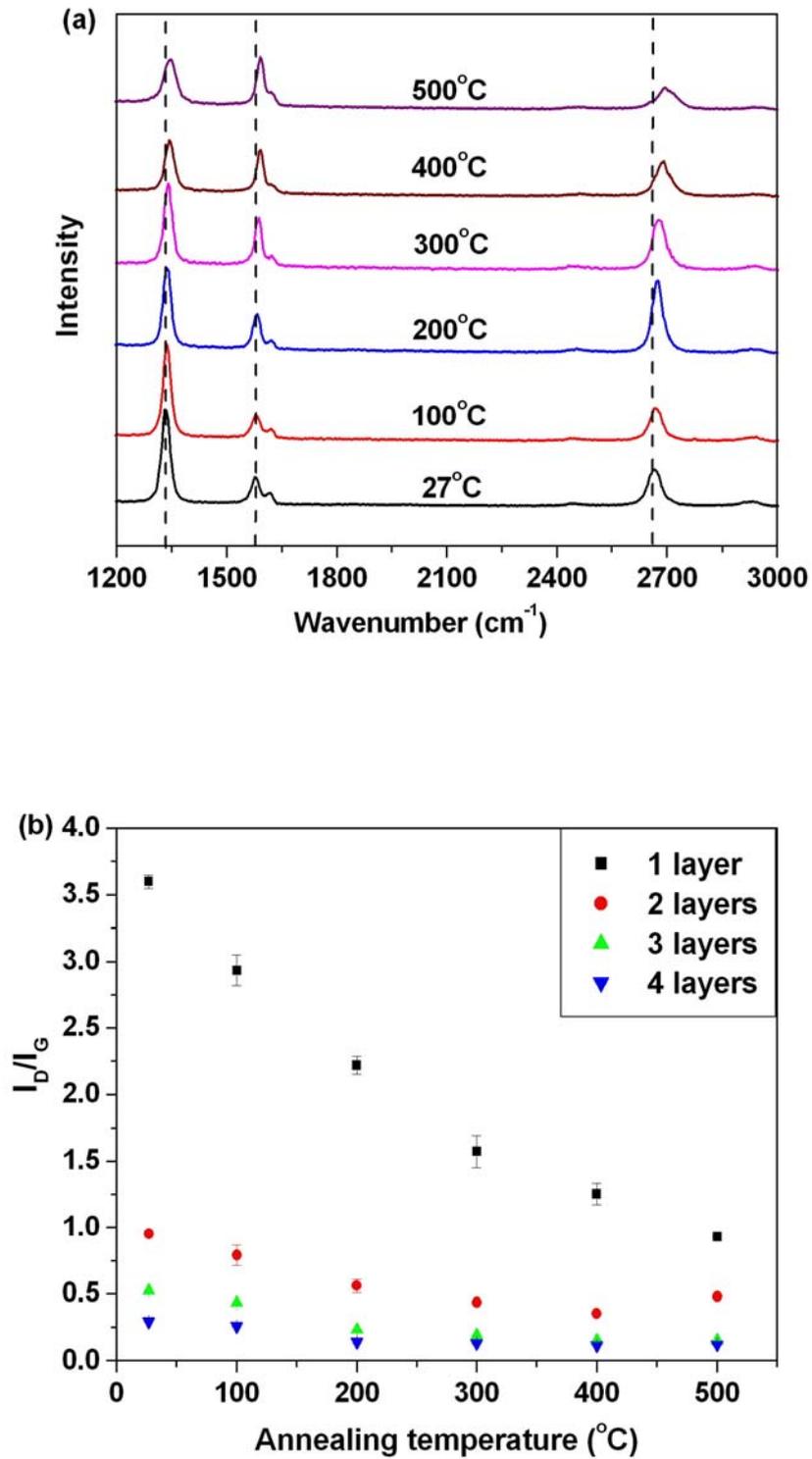

Figure 4. (a) Raman spectra of single layer graphene coated by 5 nm $SiO_2$ and annealed at different temperature. (b) The intensity ratio of D band and G band ($I_D/I_G$) of graphene sheets with one to four layers (coated with $SiO_2$) after annealing at different temperature. The $I_D/I_G$ (defects) decreased significantly upon annealing.

Fig. 5

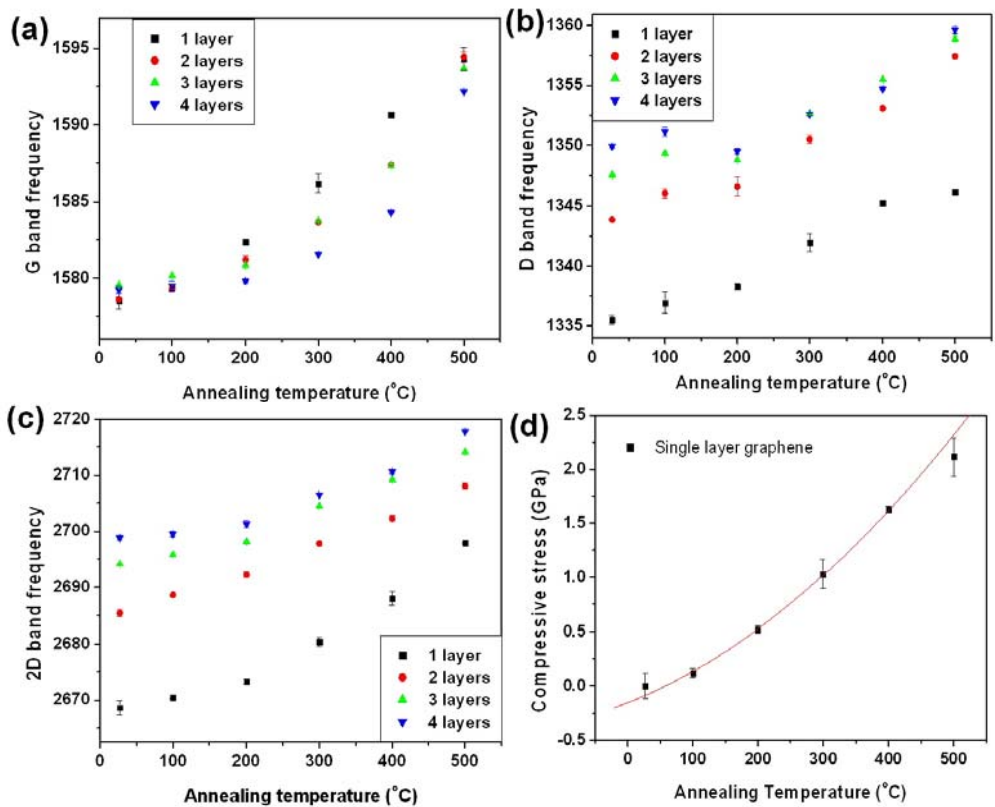

Figure 5. The Raman peak frequency of G band (a), D band (b), and 2D band (c) of graphene sheets with one to four layers (coated with $SiO_2$) after annealing at differenet temperature. Blue-shifts of all the Raman bands were observed after annealing, which were attributed to the strong compressive stress on graphene. (d) Magnitude of compressive stress on single layer graphene controlled by annealing temperature. The red line is a curve fit to the experimental data.

Fig. 6

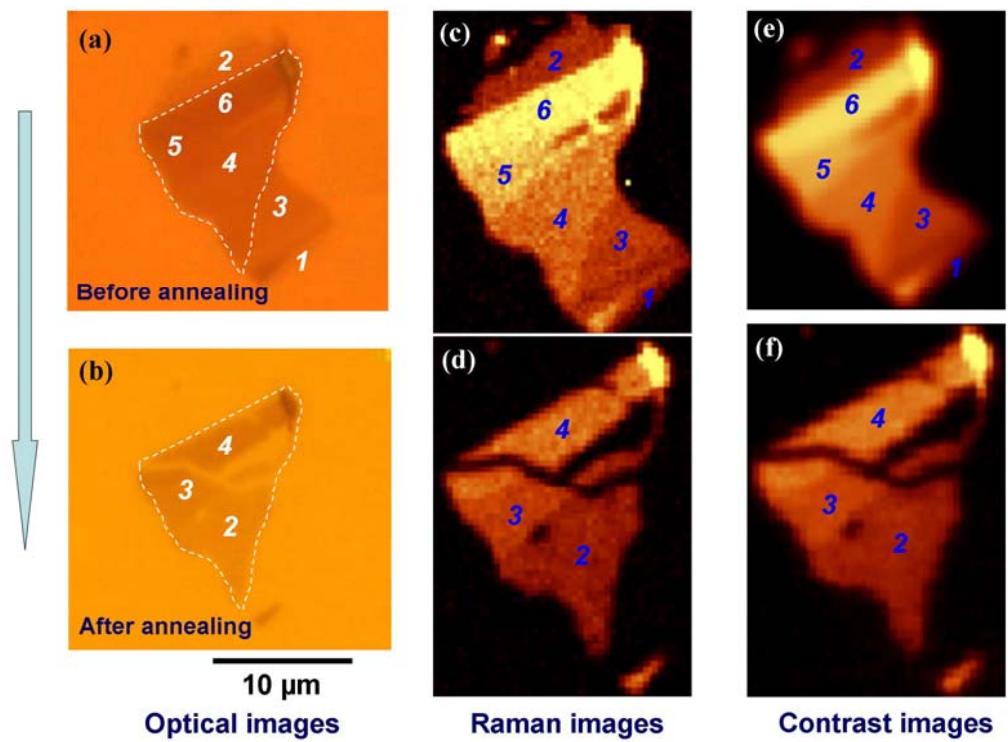

Figure 6. Optical images of a graphene sheet with one, two, three, four, and six layer regions before (a) and after (b) after annealed at 600 °C for 30 min. Raman (G band intensity) images of the same graphene before (c) and after (d) annealing. Contrast images of the same graphene before (e) and after (f) annealing. The one to three layer regions disappeared after annealing, while the four to six layer regions remained. The thicknesses of three remained regions were two, three, and four layers determined by Raman and contrast imaging.

Fig. 7

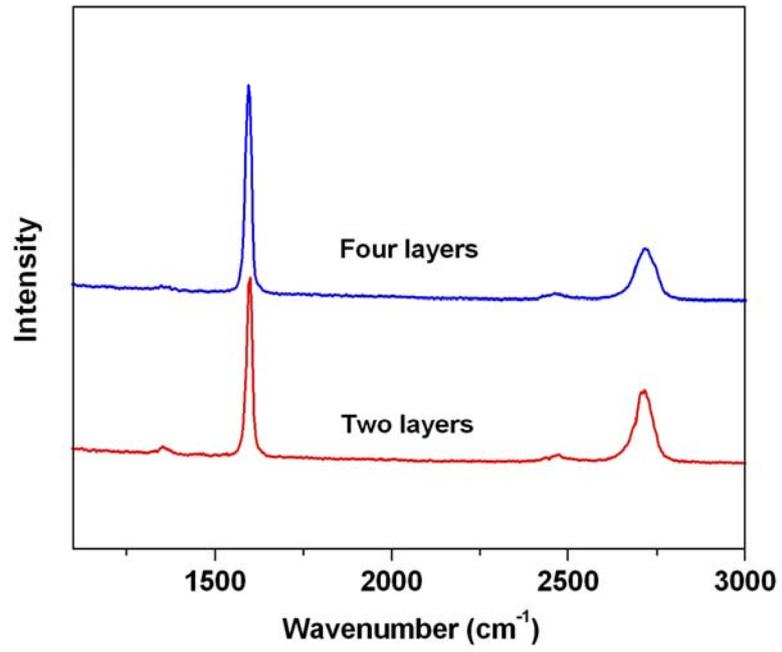

Figure 7. Raman spectra of the two and four layer graphene after thickness modification.